\documentclass[11pt,a4paper,twoside]{article} 

\usepackage{amsmath} 
\usepackage{amssymb} 
\usepackage{amsthm} 
\usepackage{graphicx}
\usepackage{psfrag}

\newcommand{\pa}[1]{{\partial#1}}

\newcommand{\h}[1]{{\hat#1}}
\newcommand{\ti}[1]{{\tilde#1}}

\newcommand{\ga}{\alpha}

\newcommand{\gga}{\gamma}
\newcommand{\gl}{\lambda}
\newcommand{\gom}{\omega}

\newcommand{\gs}{\sigma}
\newcommand{\gt}{\theta}
\newcommand{\gz}{\zeta}

\def\XXint#1#2#3{{\setbox0=\hbox{$#1{#2#3}{\int}$}
     \vcenter{\hbox{$#2#3$}}\kern-.5\wd0}}

\theoremstyle{plain} 

\newtheorem{theorem}{Theorem}[section] 
\newtheorem{proposition}[theorem]{Proposition}

\newtheorem{remark}[theorem]{Remark}

\numberwithin{equation}{section} 
\numberwithin{figure}{section} 

\begin{document} 

\title{The Generalized Dirichlet to Neumann map for the KdV equation on the half-line}
\author{P.A.~Treharne \\
School of Mathematics and Statistics \\
University of Sydney \\
Sydney 2006, Australia \\
(P.Treharne@maths.usyd.edu.au) \\
and \\
A.S.~Fokas \\
Department of Applied Mathematics and Theoretical Physics \\
University of Cambridge \\
Cambridge CB3 0WA, UK \\
(T.Fokas@damtp.cam.ac.uk)}
\date{October 2006}

\maketitle

\abstract

For the two versions of the KdV equation on the positive half-line an initial-boundary value problem is well posed if one prescribes an initial condition plus either one boundary condition if $q_{t}$ and $q_{xxx}$ have the same sign (KdVI) or two boundary conditions if $q_{t}$ and $q_{xxx}$ have opposite sign (KdVII).  Constructing the generalized Dirichlet to Neumann map for the above problems means characterizing the unknown boundary values in terms of the given initial and boundary conditions.  For example, if $\{ q(x,0),q(0,t) \}$ and $\{ q(x,0),q(0,t),q_{x}(0,t) \}$ are given for the KdVI and KdVII equations, respectively, then one must construct the unknown boundary values $\{ q_{x}(0,t),q_{xx}(0,t) \}$ and $\{ q_{xx}(0,t) \}$, respectively.  We show that this can be achieved without solving for $q(x,t)$ by analysing a certain ``global relation'' which couples the given initial and boundary conditions with the unknown boundary values, as well as with the function $\Phi^{(t)}(t,k)$, where $\Phi^{(t)}$ satisifies the $t$-part of the associated Lax pair evaluated at $x=0$.  Indeed, by employing a Gelfand--Levitan--Marchenko triangular representation for $\Phi^{(t)}$, the global relation can be solved \emph{explicitly} for the unknown boundary values in terms of the given initial and boundary conditions and the function $\Phi^{(t)}$.  This yields the unknown boundary values in terms of a nonlinear Volterra integral equation.

\section{Introduction} \label{intro-KdV}

The Korteweg--de Vries (KdV) equation
\begin{equation} \label{KdV}
q_{t} + c q_{x} - \gl q_{xxx} + 6qq_{x} = 0, \quad c = 0,1, \quad \gl = \pm 1,
\end{equation}
appears in a wide range of physical applications.  For example, it models one-dimensional small amplitude surface gravity waves propagating in a shallow channel of water.  This equation has a celebrated history in the field of integrability; in particular it was for this equation that a method was first developed, later called the inverse scattering transform method, for solving the Cauchy (initial value) problem for integrable nonlinear evolution equations \cite{GGKM1967}, \cite{FZ1994}.  In 1968 Peter Lax realised that the key to the integrability of the KdV equation was the fact that it can be expressed as the compatibility condition of two linear eigenvalue equations \cite{L1968}, now called a ``Lax pair''.  Soon thereafter it was shown that a similar Lax pair formulation exists for the nonlinear Schr\"{o}dinger equation \cite{ZS1972}, as well as for the sine-Gordon and for the modified KdV equations.  Actually, the existence of a Lax pair depending on a complex spectral parameter, denoted here by $k$, is a defining property of integrability.

The problem of solving the KdV equation on the half-line $0 < x < \infty, t>0$, was considered in \cite{ASF2002nonlin} using the general methodology introduced by one of the authors in \cite{ASF1997} and further developed in \cite{FIS2001}.  The starting point of this method is the \emph{simultaneous} spectral analysis of the two eigenvalue equations that make up the associated Lax pair.  This yields the solution of the KdV equation in terms of the solution of a matrix Riemann--Hilbert (RH) problem.  This RH problem is formulated in the complex plane of the spectral parameter, the complex $k$-plane, and its ``jump'' matrix is uniquely specified in terms of appropriate spectral functions, which are defined in terms of the initial condition $q(x,0)$ and the boundary values $q(0,t), q_{x}(0,t), q_{xx}(0,t)$.  Thus, the solution of a well posed initial-boundary value problem for the KdV equation can be expressed in terms of a matrix RH problem with \emph{known} jumps, provided that one can construct the generalized Dirichlet to Neumann map, i.e.~one can characterize the unknown boundary values in terms of the given initial and boundary conditions.  It was shown in \cite{ASF1997} that this can be achieved, in principle, by analysing the so-called \emph{global relation}, which is a simple algebraic relation coupling the spectral functions in the complex $k$-plane.  Since these functions are defined in terms of the initial condition and all the boundary values, the global relation provides an implicit characterization of the Dirichlet to Neumann map.  A breakthrough in the analysis of the global relation was announced in \cite{BdMFS2003} , where it was shown that the global relation for the nonlinear Schr\"{o}dinger equation can be solved \emph{explicitly} in terms of the given initial and boundary conditions, as well as in terms of $\Phi^{(t)}(t,k)$, the solution of the $t$-part of the associated Lax pair evaluated at $x=0$.  Following this development, the generalized Dirichlet to Neumann map for several integrable nonlinear PDEs was analysed in \cite{ASF2004dirich}, but the corresponding problem for equation \eqref{KdV} was not addressed.  This omission was due to the fact that for the explicit solution of the global relation one uses an appropriate Gelfand--Levitan--Marchenko representation for $\Phi^{(t)}(t,k)$ and, in the case of equation \eqref{KdV}, such a representation was not known until now.  In this paper, we present an appropriate Gelfand--Levitan--Marchenko representation for \eqref{KdV} and then use this representation to solve the global relation explicitly.

The paper is organized as follows.  In Section~\ref{results} we solve the global relation and present our main result.  In Section~\ref{GLM} we derive a Gelfand--Levitan--Marchenko representation for $\Phi^{(t)}(t,k)$.  In Section~\ref{D2N}, using the techniques developed in \cite{BdMFS2003}, we show that it is possible to solve the global relation for both versions of the KdV equation explicitly.  Actually, by computing certain $k$-integrals (see also \cite{Z2006}) we carry the technique of \cite{BdMFS2003} one step further and show that the relevant expressions can be simplified substantially, see Section~\ref{results}.  These results are discussed further in Section~\ref{Conclusion}.  In Appendix A we review the general methodology of \cite{ASF2002nonlin}, namely we first analyse the ``direct'' spectral problem by defining appropriate spectral functions $\{ a(k), b(k) \}$ and $\{ A(k), B(k) \}$, associated with the initial condition and with the boundary values, respectively.  We then formulate the ``inverse'' spectral problem through a matrix RH problem defined in terms of the above spectral functions.  The solution of this problem yields the solution of the KdV equation with prescribed initial and boundary data, provided that the spectral functions satisfy the global relation.  In Appendix B we discuss the generalized Dirichlet to Neumann map for the linear version of equation \eqref{KdV}.  

\begin{remark}{\rm
Equation \eqref{KdV} is written in laboratory coordinates. We note that the choice $\gl = 1$ corresponds to the KdV equation with dominant surface tension.  We also note that the usual caveat where one removes the $q_{x}$ term from equation \eqref{KdV} by changing to travelling coordinates is not available in the quarter-plane without introducing a moving boundary problem (see \cite{BSZ2002}).  If one drops the $q_{x}$ term arbitrarily (i.e.~take $c=0$ in equation \eqref{KdV}), the resulting initial-boundary value problem may be treated by a considerably simplified version of the analysis that is developed here.}
\end{remark}

\subsection{Notations and Assumptions}
We use the following notations and assumptions:
\begin{itemize}
\item
Subscripts with respect to $x$ and $t$ denote partial derivatives, for example
\[
q_{t} = \frac{\pa{q}}{\pa{t}}, \quad q_{xxx} = \frac{\pa{}^{3}{q}}{\pa{t}^3}.
\]

\item
$\overline{f(k)}$ denotes the complex conjugate of the function $f(k)$.

\item
$q(x,0) = q_{0}(x)$ denotes the initial condition.  It is assumed that the initial condition belongs to the Schwarz space on the half-line, which will be denoted by $S(\mathbb{R}^{+})$.

\item
We denote the boundary values at $x=0$ by
\begin{equation} \label{BVs}
q(0,t)=g_{0}(t), \quad q_{x}(0,t)=g_{1}(t), \quad q_{xx}(0,t)=g_{2}(t).
\end{equation}
A subset of these functions is presribed as boundary conditions.  We assume that these boundary conditions are sufficiently smooth and are compatible with $q_{0}(x)$ at $x=t=0$.

\item
$A_{11},A_{12},A_{21},A_{22}$ denote the $(11),(12),(21),(22)$ entries, respectively, of the $2\times2$ matrix $A$.

\item
$\overline{D}$ denotes the closure of the domain $D$ in the complex $k$-plane.

\item
The dispersion relation $\gom(k)$ is defined by
\begin{equation} \label{dispersion}
\gom(k) = 4\gl k^{3} + c k, \quad \gl = \pm 1, \quad c = 0,1.
\end{equation}

\end{itemize}

\section{The Main Result} \label{results}

Given the initial condition $q_{0}(x)$, define the spectral functions $a(k)$ and $b(k)$ by
\begin{equation} \label{spectral(a,b)}
a(k) = \psi_{2}(0,k), \quad b(k) = \psi_{1}(0,k), \quad \mathrm{Im}\, k \leq 0,
\end{equation}
where the vector $\big( \psi_{1}(x,k),\psi_{2}(x,k) \big)$ satisfies the following ODE, which is uniquely specified in terms of $q(x,0)$:
\begin{subequations} \label{a,b}
\begin{equation} \label{x-ODE}
\begin{split}
&\frac{\pa{\psi_{1}}}{\pa{x}} - 2ik \psi_{1} = \frac{\gl}{2ik} q_{0}(x) \big( \psi_{1} + \psi_{2} \big), \\
&\frac{\pa{\psi_{2}}}{\pa{x}} = -\frac{\gl}{2ik} q_{0}(x) \big( \psi_{1} + \psi_{2} \big),
\end{split}
\end{equation}
where
\begin{equation*}
0 < x < \infty, \quad \mathrm{Im}\, k \leq 0,
\end{equation*}
and
\begin{equation}
\lim_{x \to \infty} \psi_{1}(x,k) = 0, \quad \lim_{x \to \infty} \psi_{2}(x,k) = 1.
\end{equation}
\end{subequations}

The global relation associated with the KdV equation is given by  (see \cite{ASF2002nonlin})
\begin{equation} \label{global-scalar}
a(k)B(t,k) - b(k)A(t,k) = e^{2i\gom(k)t}c(t,k), \quad \mathrm{Im}\, k \leq 0, \quad 0 < t < \infty,
\end{equation}
where $\gom(k)$ is defined by equation \eqref{dispersion}, $c(t,k)$ is an analytic function of $k$ for $\mathrm{Im}\,k < 0$ which is of $O(1/k)$ as $k \to \infty$, and the functions $\{ A(t,k),B(t,k) \}$ are defined by
\begin{equation} \label{spectral(A,B)}
A(t,k) = \overline{\Phi_{2}(t,\bar{k})}, \quad B(t,k) = -e^{2i\gom(k)t} \Phi_{1}(t,k),
\end{equation}
where the vector $\big( \Phi_{1}(t,k),\Phi_{2}(t,k) \big)^{T}$ satisfies the following ODE, which is uniquely specified in terms of $\{ q(0,t),q_{x}(0,t),q_{xx}(0,t) \}$:
\begin{subequations} \label{A,B}
\begin{equation} \label{t-ODE}
\begin{split}
&\frac{\pa{\Phi_{1}}}{\pa{t}} + 2i\gom(k) \Phi_{1} = 2ikg_{0}(t) \Phi_{2} + g_{1}(t) \Phi_{2} + \frac{1}{2ik} \big(g_{2}(t) - \gl c g_{0}(t) - 2\gl g_{0}(t)^{2} \big) \big( \Phi_{1} + \Phi_{2} \big), \\
&\frac{\pa{\Phi_{2}}}{\pa{t}} = -2ikg_{0}(t) \Phi_{1} + g_{1}(t) \Phi_{1} - \frac{1}{2ik} \big(g_{2}(t) - \gl c g_{0}(t) - 2\gl g_{0}(t)^{2} \big) \big( \Phi_{1} + \Phi_{2} \big),
\end{split}
\end{equation}
where
\begin{equation}
0 < t < \infty, \quad k \in \mathbb{C},
\end{equation}
and
\begin{equation}
\Phi_{1}(0,k) = 0, \quad \Phi_{2}(0,k) = 1.
\end{equation}
\end{subequations}

In the remainder of this section we give our main results and show that, for both versions of the KdV equation, the global relation \eqref{global-scalar} can be solved \emph{explicitly} for the unknown boundary values, ie.~we give the generalized Dirichlet to Neumann map for equation \eqref{KdV}.  These results will be proved in Sections~\ref{GLM} and \ref{D2N} by using a Gelfand--Levitan--Marchenko triangular representation for the spectral functions $A(t,k)$ and $B(t,k)$.

\subsection{The Gelfand--Levitan--Marchenko representation of $\Phi_{1}$ and $\Phi_{2}$}

\begin{proposition} \label{proposition}
The vector $\big( \Phi_{1}(t,k),\Phi_{2}(t,k) \big)^{T}$ has the following Gelfand--Levitan--Marchenko representation
\begin{equation} \label{GLM-1}
\begin{pmatrix}
\Phi_{1} \\
\Phi_{2}
\end{pmatrix} = 
\begin{pmatrix}
0 \\
1
\end{pmatrix} + \int_{-t}^{t}
\begin{pmatrix}
F_{1}(t,s,k) \\
F_{2}(t,s,k)
\end{pmatrix}
e^{i\gom(k) (t - s) \gs_{3}} ds,
\end{equation}
where 
\begin{align*}
F_{1}(t,s,k) &= -kB_{1} + C_{1} + \frac{1}{4\gl} g_{0}(t)A_{1} - \frac{1}{k} \big[ D_{1} - \frac{1}{4\gl} g_{0}(t) B_{2} - \frac{1}{8i\gl} g_{1}(t) A_{1} \big] \\
F_{2}(t,s,k) &= k^{2} A_{1} + k B_{2} + C_{2} + \frac{1}{k} \big[ D_{1} - \frac{1}{4\gl} g_{0}(t) B_{2} - \frac{1}{8i\gl} g_{1}(t) A_{1} \big],
\end{align*}
and the six scalar functions $\{ A_{1}(t,s), B_{1}(t,s), B_{2}(t,s), C_{1}(t,s), C_{2}(t,s), D_{1}(t,s) \}$ satisfy the following well-posed characteristic-value problem:
\begin{subequations} \label{charac-value-DEs2}
\begin{align}
&(\pa{}_{t} + \pa{}_{s}) A_{1} = 2ig_{0} B_{1} + 8i\gl D_{1} \\
&(\pa{}_{t} - \pa{}_{s}) B_{1} = \big( G_{2} - \frac{1}{2i\gl} g_{0} \big) A_{1} - g_{1} B_{2} - 2ig_{0} C_{2} \\
&(\pa{}_{t} + \pa{}_{s}) B_{2} = \big( G_{2} + \frac{1}{2i\gl} g_{0}^{2} \big) A_{1} - g_{1} B_{1} - 2ig_{0} C_{1} \\
&(\pa{}_{t} - \pa{}_{s}) C_{1} = - \frac{1}{4\gl} \left( \dot{g}_{0} + g_{0} g_{1} + g_{1} \right) A_{1} + \big( G_{2} + \frac{1}{2i\gl} g_{0}^{2} \big) B_{1} \notag \\
&\hspace{55mm} + \big( \frac{1}{2i\gl} g_{0} - G_{2} + \frac{1}{2i\gl} g_{0}^{2} \big) B_{2} + g_{1} C_{2} \\
&(\pa{}_{t} + \pa{}_{s}) C_{2} = - G_{2} B_{1} + \big( G_{2} + \frac{1}{2i\gl} g_{0}^{2} \big) B_{2} + g_{1} C_{1} + \big( 2i + 2ig_{0} \big) D_{1} \\
&(\pa{}_{t} - \pa{}_{s}) D_{1} = \frac{1}{8i\gl} \big( \dot{g}_{1} + g_{1}^{2} - 2 G_{0}G_{2} + \frac{1}{\gl} g_{0}^{3} \big) A_{1} + \frac{1}{2\gl} g_{0} \pa{}_{s} B_{2} \notag \\
&\hspace{35mm} + \frac{1}{4\gl} \big( \dot{g}_{0} + g_{0}g_{1} \big) B_{2} + \big( G_{2} + \frac{1}{2i\gl} g_{0}^{2} \big) C_{1} + G_{2} C_{2},
\end{align}
\end{subequations}
where $\dot{g}_{0} = \frac{d}{dt}g_{0}$, $\dot{g}_{1} = \frac{d}{dt}g_{1}$ and $G_{2} = \frac{i}{2}(g_{2} - \gl c g_{0} - 2\gl g_{0}^{2})$, with boundary conditions
\begin{equation} \label{charac-value-BCs2+}
B_{1}(t,t) = -ig_{0}(t), \quad C_{1}(t,t) = \frac{1}{2} g_{1}(t), \quad D_{1}(t,t) = \frac{i}{4} \big(g_{2}(t) - \gl c g_{0}(t) - 2\gl g_{0}(t)^2 \big),
\end{equation}
and 
\begin{equation} \label{charac-value-BCs2-}
A_{1}(t,-t) = B_{2}(t,-t) = C_{2}(t,-t) = 0.
\end{equation}
\end{proposition}

\noindent
\textbf{Proof.}
The proof of this Proposition will be given below in Section~\ref{GLM}.

\subsection{The generalized Dirichlet to Neumann map for KdVI}

\begin{figure}
\psfrag{a}{$D_{1}$}
\psfrag{b}{$\pa{D}$}
\begin{center} 
\includegraphics{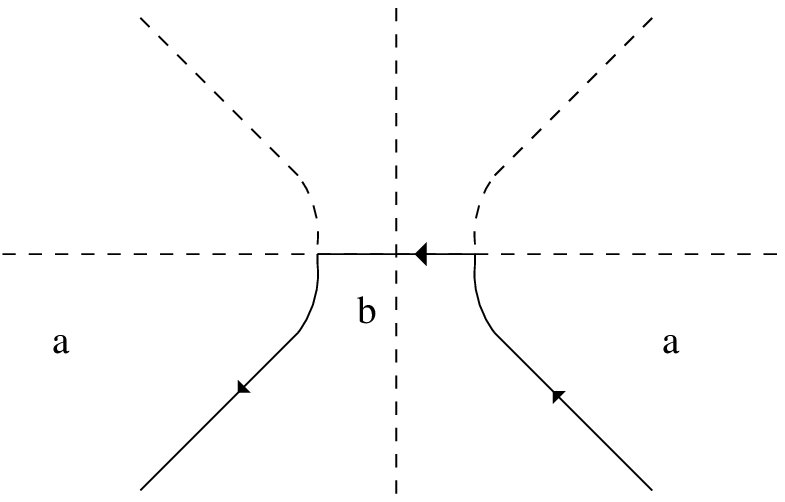}
\caption{The oriented boundary $\pa{D}$ for the case $\gl = -1$.} \label{boundary-D1*}
\end{center} 
\end{figure}

For the KdV equation \eqref{KdV} with $\gl = -1$, an initial-boundary value problem is well posed if we specify the initial condition $q(x,0)=q_{0}(x)$ together with a \emph{single} boundary condition, say $q(0,t)=g_{0}(t)$.  In this case the analysis of the global relation yields two algebraic relations which can be solved explicitly to determine the unknown boundary values $g_{1}(t)$ and $g_{2}(t)$.  We will show that
\begin{subequations} \label{Isolutions}
\begin{multline} \label{Ig1}
g_{1}(t) = \frac{1}{2} g_{0}(t) A_{1}(t,t) - 3^{-1/3} \int_{0}^{t} \frac{Ai(\gz)}{(t-\tau)^{1/3}} \Big[ 3i\frac{\pa{B}_{1}}{\pa{\tau}} - 4B_{1} \Big] (t,2\tau-t) d\tau \\
- \frac{1}{\pi} \int_{\pa{D}} \frac{\gom'(k)e^{-2i\gom(k)t}}{p_{+} - p_{-}} \Big[ p_{+} \frac{b(p_{+})}{a(p_{+})} A(t,p_{+}) - p_{-} \frac{b(p_{-})}{a(p_{-})} A(t,p_{-}) \Big] dk,
\end{multline}
\begin{multline} \label{Ig2}
g_{2}(t) = - 2g_{0}(t)^{2} + i g_{0}(t) B_{2}(t,t) - \frac{1}{2} g_{1}(t) A_{1}(t,t)  \\
- 3^{-2/3} \int_{0}^{t} \frac{Ai'(\gz)}{(t-\tau)^{2/3}} \Big[ \frac{3i}{2}\frac{\pa{B}_{1}}{\pa{\tau}} + 2B_{1} \Big] (t,2\tau-t) d\tau \\
+ \frac{2i}{\pi} \int_{\pa{D}} \gom'(k)e^{-2i\gom(k)t} \frac{p_{+}p_{-}}{p_{+} - p_{-}} \Big[ \frac{b(p_{+})}{a(p_{+})} A(t,p_{+}) - \frac{b(p_{-})}{a(p_{-})} A(t,p_{-}) \Big] dk,
\end{multline}
\end{subequations}
where $A(t,k)$ is defined by the first equation in \eqref{spectral(A,B)}, $Ai(\gz)$ is a solution of Airy's equation given by
\begin{equation} \label{Airy}
Ai(\gz) = \frac{1}{2\pi} \int_{\pa{D}} e^{i(\gz k + k^{3}/3)} dk, \quad \gz = -3^{-1/3}(t - \tau)^{2/3},
\end{equation}
the oriented contour $\pa{D}$ is shown in Figure~\ref{boundary-D1*}, $\gom(k)$ is defined by equation \eqref{dispersion} with $\gl = -1$ and $\gom'(k)$ denotes the derivative of $\gom(k)$, and $p_{\pm}$ are the two non-trivial roots of $\gom(p)=\gom(k)$
\begin{equation} \label{nu*}
p_{\pm}(k) = -\frac{k}{2} \pm \frac{i}{2} (3k^2 - c)^{1/2}.
\end{equation}

\subsection{The generalized Dirichlet to Neumann map for KdVII}

\begin{figure}
\psfrag{a}{$D_{1}$}
\psfrag{b}{$\pa{D}$}
\begin{center} 
\includegraphics{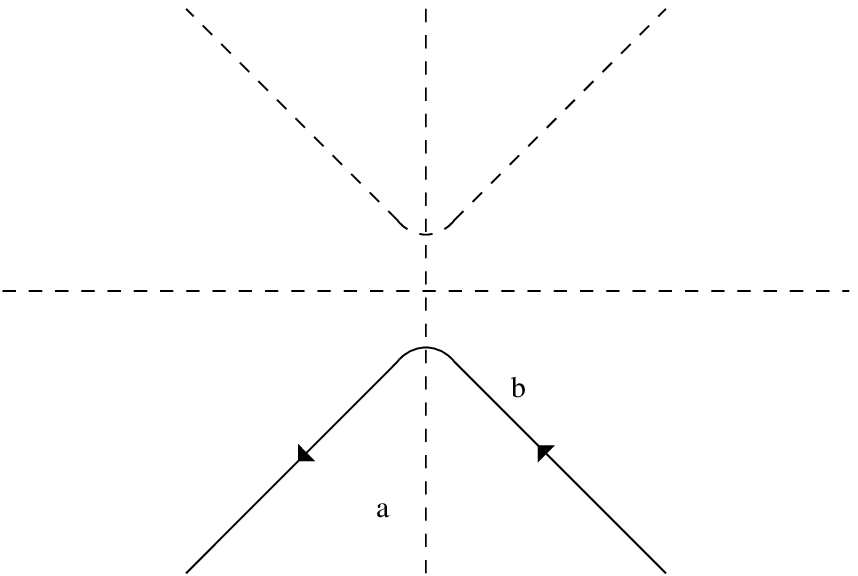}
\caption{The oriented boundary $\pa{D}$ for the case $\gl = 1$.} \label{boundary-D1}
\end{center} 
\end{figure}

For the KdV equation \eqref{KdV} with $\gl = 1$, the initial-boundary value problem is well posed if we specify the initial condition $q(x,0)=q_{0}(x)$ together with \emph{two} independent boundary conditions.  To simplify the analysis, we consider the case where $q(0,t)=g_{0}(t)$ and $q_{x}(0,t)=g_{1}(t)$ are given and solve the global relation explicitly for the unknown boundary value $g_{2}(t)$.  We will show that
\begin{multline} \label{IIg2}
g_{2}(t) = cg_{0}(t) + 2g_{0}(t)^{2} - i g_{0}(t) B_{2}(t,t) - \frac{1}{2} g_{1}(t) A_{1}(t,t)  \\
+ 3^{-2/3} \int_{0}^{t} \frac{Ai'(\gz)}{(t-\tau)^{2/3}} \Big[ \frac{3i}{2} \frac{\pa{B}_{1}}{\pa{\tau}} + 2B_{1} \Big] (t,2\tau-t) d\tau \\
- 3^{-1/3} \int_{0}^{t} \frac{Ai(\gz)}{(t-\tau)^{1/3}} \Big[ 6\frac{\pa{C}_{1}}{\pa{\tau}} + 8iC_{1} + g_{0}(t) \big( \frac{3}{2} \frac{\pa{A}_{1}}{\pa{\tau}} + 2iA_{1}) \Big] (t,2\tau-t) d\tau \\
- \frac{2i}{\pi} \int_{\pa{D}} k \gom'(k) e^{-2i\gom(k)t} \frac{b(k)}{a(k)} A(t,k) dk,
\end{multline}
where $Ai(\gz)$ is given by \eqref{Airy} but with $\gz = 3^{-1/3}(t - \tau)^{2/3}$, $\gom(k)$ is defined by equation \eqref{dispersion} with $\gl = 1$ and $\gom'(k)$ denotes the derivative of $\gom(k)$, and the contour $\pa{D}$, shown in Figure~\ref{boundary-D1}, is the oriented boundary of the domain 
\begin{equation*}
D_1 = \{k \in \mathbb{C}, \mathrm{Im}\; k < 0 \cap \mathrm{Im}\; \gom(k) > 0 \}.
\end{equation*}

\begin{remark} {\rm
We note that the explicit solution of the global relation given in \eqref{Isolutions} and \eqref{IIg2} is expressed in terms of the given initial and boundary conditions and in terms of the kernel functions that appear in the Gelfand--Levitan--Marchenko representation of $\Phi_{1}(t,k)$ and $\Phi_{2}(t,k)$.  We note that that it is also possible to express the Dirichlet to Neumann map in terms of $\Phi_{1}(t,k)$ and $\Phi_{2}(t,k)$ only (i.e.~without reference to the kernel functions), although the resulting formulas are more complicated than those given here (see \cite{ASF2004dirich}).
}
\end{remark}

\begin{remark} {\rm
If we take $q(x,0)=0$ as our initial condition then the spectral function $b(k)$ defined in \eqref{spectral(a,b)} is zero and so the formulas given above in \eqref{Isolutions} and \eqref{IIg2} simplify considerably (the $k$-integrals in these formulas are now zero also).  Furthermore, for the special case where we take $c = 0$ in equation \eqref{KdV}, i.e.~where the $q_{x}$ term is absent, the formulas \eqref{Isolutions} and \eqref{IIg2} simplify further in that, instead of the Airy function \eqref{Airy}, they now involve the gamma function $\Gamma(\ga)$ where $\ga$ is independent of $\tau$.  Thus, for instance, if $q(x,0)=0$ and we take $c=0$ and $\gl=-1$ in equation \eqref{KdV}, then for KdVI the unknown boundary values $g_{1}(t)$ and $g_{2}(t)$ can be expressed in terms of $q(0,t)=g_{0}(t)$ by the following formulas:
\begin{subequations} 
\begin{align}
g_{1}(t) &= \frac{1}{2} g_{0}(t) A_{1}(t,t) + \frac{1}{6i\pi} \Gamma(\tfrac{1}{3}) \int_{0}^{t} \frac{1}{(t-\tau)^{1/3}} \Big[ 3i\frac{\pa{B}_{1}}{\pa{\tau}} - 4B_{1} \Big] (t,2\tau-t) d\tau, \\
g_{2}(t) &= - 2g_{0}(t)^{2} + i g_{0}(t) B_{2}(t,t) - \frac{1}{2} g_{1}(t) A_{1}(t,t) \notag \\
&\hspace{5mm} - \frac{1}{6i\pi} \Gamma(\tfrac{2}{3}) \int_{0}^{t} \frac{1}{(t-\tau)^{2/3}} \Big[ \frac{3i}{2}\frac{\pa{B}_{1}}{\pa{\tau}} + 2B_{1} \Big] (t,2\tau-t) d\tau,
\end{align}
\end{subequations}
where $\{ A_{1}, B_{1}, B_{2}, C_{1}, C_{2}, D_{1} \}$ satisfy the hyperbolic system \eqref{charac-value-DEs2}--\eqref{charac-value-BCs2-}.
}
\end{remark}

\begin{remark} {\rm
It is shown in Appendix B that, in the linear limit of small $q_{0}(x)$, the expressions in \eqref{Isolutions} and \eqref{IIg2} reduce to the \emph{same} representations that are obtained by solving the global relation associated with the linearized KdV equation 
\begin{equation} \label{KdVlinear}
q_{t} + c q_{x} - \gl q_{xxx} = 0.
\end{equation}
}
\end{remark}

\section{The Gelfand--Levitan--Marchenko representation of $\Phi^{(t)}(t,k)$} \label{GLM}

In this section we verify Proposition~\ref{proposition} directly.  Let the matrix $\Phi^{(t)}(t,k)$ be defined in terms of $(\Phi_{1},\Phi_{2})$ by
\begin{equation} \label{Phi-t}
\Phi^{(t)}(t,k) =
\begin{pmatrix}
\overline{\Phi_{2}(t,\bar{k})} & \Phi_{1}(t,k) \\
\overline{\Phi_{1}(t,\bar{k})} & \Phi_{2}(t,k)
\end{pmatrix}.
\end{equation}
Then equations \eqref{A,B} imply that this matrix satisfies the equation
\begin{subequations} \label{differential-Phi}
\begin{equation} \label{t-equation}
\pa{}_{t}\Phi^{(t)}(t,k) + i\gom(k) [\gs_{3},\Phi^{(t)}(t,k)] = \left[ kQ_{0}(t) + Q_{1}(t) + \frac{1}{k}Q_{2}(t) \right] \Phi^{(t)}(t,k),
\end{equation}
where
\begin{equation} \label{Q0-Q2}
Q_{0}(t) = G_{0}(t)
\begin{pmatrix}
0 & -1 \\
1 & 0
\end{pmatrix}, \quad
Q_{1}(t) = g_{1}(t)
\begin{pmatrix}
0 & 1 \\
1 & 0
\end{pmatrix}, \quad
Q_{2}(t) = G_{2}(t)
\begin{pmatrix}
-1 & -1 \\
1 & 1
\end{pmatrix},
\end{equation}
and
\begin{equation} \label{G0-G2}
G_{0}(t) = -2ig_{0}(t), \quad G_{2}(t) = \frac{i}{2} \big(g_{2}(t) - \gl c g_{0}(t) - 2\gl g_{0}(t)^2 \big).
\end{equation}
\end{subequations}
Substituting the representation (cf.~\cite{BdMFS2003} and \cite{ASF2004dirich})
\begin{equation} \label{GLM-Phi}
\Phi^{(t)}(t,k) = I + \int_{-t}^{t} \left( k^{2} \ti{A}(t,s) + k \ti{B}(t,s) + \ti{C}(t,s) + \frac{1}{k} \ti{D}(t,s) \right) e^{i\gom(k) (t - s) \gs_{3}} ds,
\end{equation}
into equation \eqref{t-equation} we obtain an integral equation involving the matrix-valued functions $\ti{A}, \ti{B}, \ti{C}, \ti{D}$.  By multiplying by $k^{2}$ the following identity (obtained by integration by parts)
\begin{multline} \label{ident1}
\int_{-t}^{t} k^{3} F(s)e^{i\gom(k) (t - s) \gs_{3}} ds = \frac{i\gl}{4} \left( F(t) \gs_{3} - F(-t)\gs_{3} e^{2i\gom(k)t\gs_{3}} \right) \\
-\frac{\gl}{4} \int_{-t}^{t} \big( k F(s) + iF_{s}(s)\gs_{3} \big) e^{i\gom(k) (t - s) \gs_{3}} ds,
\end{multline}
we can eliminate the integral terms involving $k^{5}$.  Then, using the identity obtained from multiplying \eqref{ident1} by $k$, as well as \eqref{ident1} itself, we can eliminate the integral terms involving $k^{4}$ and $k^{3}$.  Finally, by using the identity obtained from multiplying \eqref{ident1} by $1/k$ we can eliminate the integral terms involving $k^{2}$.  The remaining terms involve $k$, $1$, $1/k$, and $1/k^2$.   The latter term involves the product $Q_{2}\ti{D}$, which is equal to zero if the matrix $\ti{D}$ has the following form
\begin{equation} \label{Dmat}
\ti{D}(t,s) = \ti{D}_{1}(t,s)
\begin{pmatrix}
-1 & -1 \\
1 & 1
\end{pmatrix}.
\end{equation}
Equating to zero the coefficients of the terms involving $k$, $1$, $1/k$ we obtain the following equations:
\begin{subequations} \label{DEs}
\begin{align}
&\ti{B}_{t} + \gs_{3}\ti{B}_{s}\gs_{3} = Q_{0}\ti{C} + Q_{1}\ti{B} + Q_{2}\ti{A} - \frac{1}{4\gl} Q_{0}\ti{A} \\
&\ti{C}_{t} + \gs_{3}\ti{C}_{s}\gs_{3} - \frac{1}{4\gl} \big( \ti{A}_{t} + \gs_{3} \ti{A}_{s} \gs_{3} \big) = Q_{0}\ti{D} + Q_{1}\ti{C} + Q_{2}\ti{B} \notag \\
&\hspace{60mm} - \frac{1}{4\gl} \big( Q_{0} \ti{B} + Q_{1} \ti{A} \big) + \frac{1}{4i\gl} Q_{0} \ti{A}_{s} \gs_{3} \\
&\ti{D}_{t} + \gs_{3}\ti{D}_{s}\gs_{3} + \frac{1}{4i\gl} \big( \ti{A}_{ts}\gs_{3} + \gs_{3}\ti{A}_{ss}\big) = Q_{1}\ti{D} + Q_{2}\ti{C} + \frac{1}{4i\gl} \big( Q_{0} \ti{B}_{s} + Q_{1} \ti{A}_{s} \big) \gs_{3},
\end{align}
\end{subequations}
Equating the coefficients of the boundary terms involving $k^2, k, 1, 1/k$ we find the following boundary conditions:
\begin{subequations} \label{BCs-1}
\begin{align}
\ti{A}(t,t) - \gs_{3}\ti{A}(t,t)\gs_{3} &= 0 \\
\ti{B}(t,t) - \gs_{3}\ti{B}(t,t)\gs_{3} &= Q_{0} \\
\ti{C}(t,t) - \gs_{3}\ti{C}(t,t)\gs_{3} &= Q_{1} - \frac{1}{4i\gl}Q_{0}\ti{A}(t,t)\gs_{3} \label{C+} \\
\ti{D}(t,t) - \gs_{3}\ti{D}(t,t)\gs_{3} - \frac{1}{4i\gl} \big( &\ti{A}_{t}(t,t)\gs_{3} + \gs_{3}\ti{A}_{s}(t,t)\big) \notag \\
&= Q_{2} - \frac{1}{4i\gl} \big( Q_{0}\ti{B}(t,t) + Q_{1}\ti{A}(t,t) \big) \gs_{3}; \label{D+}
\end{align}
\end{subequations}
and
\begin{subequations} \label{BCs-2}
\begin{align}
\ti{A}(t,-t) + \gs_{3}\ti{A}(t,-t)\gs_{3} &= 0 \\
\ti{B}(t,-t) + \gs_{3}\ti{B}(t,-t)\gs_{3} &= 0 \\
\ti{C}(t,-t) + \gs_{3}\ti{C}(t,-t)\gs_{3} &= \frac{1}{4i\gl}Q_{0}\ti{A}(t,-t)\gs_{3} \label{C-} \\
\ti{D}(t,-t) + \gs_{3}\ti{D}(t,-t)\gs_{3} + \frac{1}{4i\gl} \big( &\ti{A}_{t}(t,-t)\gs_{3} + \gs_{3}\ti{A}_{s}(t,-t)\big) \notag \\
&= \frac{1}{4i\gl} \big( Q_{0}\ti{B}(t,-t) + Q_{1}\ti{A}(t,-t) \big) \gs_{3}. \label{D-}
\end{align}
\end{subequations}
We choose the matrices $\ti{A}, \ti{B}, \ti{C}$ in the following form,
\begin{equation}  \label{matrices} 
\ti{A}(t,s) = A_{1}(t,s)
\begin{pmatrix} 
1 & 0 \\ 
0 & 1
\end{pmatrix}, \quad
\ti{B}(t,s) = 
\begin{pmatrix} 
-B_{2} & -B_{1} \\ 
B_{1} & B_{2}
\end{pmatrix}, \quad
\ti{C}(t,s) = 
\begin{pmatrix} 
C_{2} & \ti{C}_{1} \\ 
\ti{C}_{1} & C_{2}
\end{pmatrix}.
\end{equation}
Substituting \eqref{Dmat} and \eqref{matrices} into equations \eqref{DEs} we find the following differential equations:
\begin{subequations} \label{charac-value-DEs}
\begin{align}
&(\pa{}_{t} - \pa{}_{s}) B_{1} = \big( G_{2} - \frac{1}{4\gl} G_{0} \big) A_{1} - g_{1} B_{2} + G_{0} C_{2} \label{CVa} \\
&(\pa{}_{t} + \pa{}_{s}) B_{2} = G_{2} A_{1} - g_{1} B_{1} + G_{0} \ti{C}_{1} \\
&(\pa{}_{t} - \pa{}_{s}) \ti{C}_{1} = \frac{1}{4i\gl} G_{0} \pa{}_{s} A_{1} - \frac{1}{4\gl} g_{1} A_{1} + G_{2} B_{1} + \big( \frac{1}{4\gl} G_{0} - G_{2} \big) B_{2} \notag \\
&\hspace{30mm} + g_{1} C_{2} - G_{0} \ti{D}_{1} \\
&(\pa{}_{t} + \pa{}_{s}) C_{2} = \frac{1}{4\gl} \big( \pa{}_{t} + \pa{}_{s} \big) A_{1} + \big( \frac{1}{4\gl} G_{0} - G_{2} \big) B_{1} + G_{2} B_{2} + g_{1} \ti{C}_{1} - G_{0} \ti{D}_{1} \label{CVd} \\
&(\pa{}_{t} - \pa{}_{s}) \ti{D}_{1} = \frac{1}{4i\gl} g_{1} \pa{}_{s} A_{1} - \frac{1}{4i\gl} G_{0} \pa{}_{s} B_{2} + G_{2} \big( \ti{C}_{1} + C_{2} \big)  - g_{1} \ti{D}_{1} \label{CVe} \\
&(\pa{}_{t} + \pa{}_{s}) \ti{D}_{1} = \frac{1}{4i\gl} (\pa{}_{t} + \pa{}_{s}) \pa{}_{s} A_{1} + \frac{1}{4i\gl} G_{0} \pa{}_{s} B_{1} + G_{2} \big( \ti{C}_{1} + C_{2} \big) - g_{1} \ti{D}_{1} \label{CVf}
\end{align}
\end{subequations}
Furthermore, equations \eqref{BCs-1} and \eqref{BCs-2} yield the following boundary conditions:
\begin{subequations} \label{charac-value-BCs+}
\begin{align}
&2B_{1}(t,t) = G_{0}(t) \label{BCa} \\
&2\ti{C}_{1}(t,t) = g_{1}(t) - \frac{1}{4i\gl} G_{0}(t) A_{1}(t,t) \label{BCb} \\
&2\ti{D}_{1}(t,t) = G_{2}(t) - \frac{1}{4i\gl} g_{1}(t) A_{1}(t,t) + \frac{1}{4i\gl} G_{0}(t) B_{2}(t,t) \label{BCc} \\
&-\frac{1}{4i\gl} (\pa{}_{t} + \pa{}_{s}) A_{1}(t,t) = - G_{2} + \frac{1}{4i\gl} G_{0}(t) B_{1}(t,t), \label{BCd}
\end{align}
\end{subequations}
and
\begin{subequations} \label{charac-value-BCs-}
\begin{align}
&A_{1}(t,-t) = 0 \label{BCa-} \\
&B_{2}(t,-t) = 0 \label{BCb-} \\
&C_{2}(t,-t) = 0 \label{BCc-} \\ 
&2\ti{D}_{1}(t,-t) - \frac{1}{4i\gl} (\pa{}_{t} + \pa{}_{s}) A_{1}(t,-t) = \frac{1}{4i\gl} G_{0} B_{1}(t,-t) \label{BCd-}\\
&0 = - \frac{1}{4i\gl} g_{1} A_{1}(t,-t) + \frac{1}{4i\gl} G_{0} B_{2}(t,-t). \label{BCe-}
\end{align}
\end{subequations}
By subtracting \eqref{CVe} from \eqref{CVf}, we find the equation
\begin{equation} \tag{3.10f$'$} \label{CVf'}
\Big\{ \big( \pa{}_{t} + \pa{}_{s} \big) A_{1} - g_{1} A_{1} - G_{0} \big( B_{1} + B_{2} \big) - 8i\gl \ti{D}_{1} \Big\}_{s} = 0.
\end{equation}
Similarly, by adding \eqref{BCc} and \eqref{BCd}, we find the boundary condition
\begin{equation} \tag{3.11d$'$} \label{BCd'}
\big( \pa{}_{t} + \pa{}_{s} \big) A_{1}(t,t) - g_{1}(t) A_{1}(t,t) + G_{0}(t) \big( B_{1}(t,t) + B_{2}(t,t) \big) - 8i\gl \ti{D}_{1}(t,t) = 0.
\end{equation}

In summary, we have derived the following result: Let $\Phi^{(t)}(t,k)$ be expressed in terms of the matrices $\{ \ti{A}, \ti{B}, \ti{C}, \ti{D} \}$ by equation \eqref{GLM-Phi}.  Let these matrices have the form given by equations \eqref{Dmat} and \eqref{matrices}.  Then $\Phi^{(t)}(t,k)$ satisfies equation \eqref{differential-Phi} provided that the scalar functions $\{ A_{1}, B_{1}, B_{2}, \ti{C}_{1}, C_{2}, \ti{D}_{1} \}$ satisfy the differential equations \eqref{CVa}--\eqref{CVe}, \eqref{CVf'} together with the boundary conditions \eqref{BCa}--\eqref{BCc}, \eqref{BCd'} at $s = t$, and the boundary conditions \eqref{charac-value-BCs-} at $s = -t$.

We will now show that the above differential equations and boundary conditions are satisfied provided that the six functions $\{ A_{1}, B_{1}, B_{2}, \ti{C}_{1}, C_{2}, \ti{D}_{1} \}$ satisfy the differential equations \eqref{CVa}--\eqref{CVe}, the differential equation
\begin{equation} \label{CVf''}
(\pa{}_{t} + \pa{}_{s}) A_{1} = g_{1} A_{1} - G_{0} \big( B_{1} + B_{2} \big) + 8i\gl \ti{D}_{1}, \tag{3.10f$''$}
\end{equation}
as well as the boundary conditions \eqref{BCa}--\eqref{BCc} and \eqref{BCa-}--\eqref{BCc-}.  Indeed, if equation \eqref{CVf''} is valid then equation \eqref{CVf'} is also valid.  Furthermore, evaluating equation \eqref{CVf'} at $s=t$, it follows that equation \eqref{BCd'} is also valid.  In addition, evaluating \eqref{CVf''} at $s=-t$ and dividing the resulting equation by $1/4i\gl$ we find
\begin{equation*}
2\ti{D}_{1}(t,-t) - \frac{1}{4i\gl} (\pa{}_{t} + \pa{}_{s}) A_{1}(t,-t) = \frac{1}{4i\gl} G_{0} \big( B_{1}(t,-t) + B_{2}(t,-t) \big) - \frac{1}{4i\gl} g_{1} A_{1}(t,-t),
\end{equation*}
which, taking into consideration equations \eqref{BCa-} and \eqref{BCb-}, yields equation \eqref{BCd-}.  Finally, equations \eqref{BCa-} and \eqref{BCb-} imply that equation \eqref{BCe-} is identically satisfied.

It is convenient to replace $(\pa{}_{t} + \pa{}_{s}) A_{1}$ in the RHS of \eqref{CVd} by the RHS of \eqref{CVf''}; thus equation \eqref{CVd} can be replaced by the equation
\begin{equation} \label{CVd'}
(\pa{}_{t} + \pa{}_{s}) C_{2} = \frac{1}{4\gl} g_{1} A_{1} - G_{2} B_{1} + \big( G_{2} - \frac{G_{0}}{4\gl} \big) B_{2} + g_{1} \ti{C}_{1} + (2i - G_{0}) \ti{D}_{1}. \tag{3.10d$'$}
\end{equation}

Equations \eqref{BCb} and \eqref{BCd'} suggest the following change of variables
\begin{align*}
\ti{C}_{1}(t,s) &= C_{1}(t,s) - \frac{1}{8i\gl} G_{0}(t) A_{1}(t,s) \\
\ti{D}_{1}(t,s) &= D_{1}(t,s) - \frac{1}{8i\gl} g_{1}(t) A_{1}(t,s) + \frac{1}{8i\gl} G_{0}(t) B_{2}(t,s).
\end{align*}
This yields the required result.

\subsection{The spectral functions $A(t,k), B(t,k)$}

Equation \eqref{Phi-t}, together with equation \eqref{differential-Phi} and the integral representation for $\Phi^{(t)}(t,k)$ in equation \eqref{GLM-Phi}, imply the following expression for $\Phi_{1}$:
\begin{align*}
\Phi_{1}(t,k) &= e^{-i\gom(k)t} \int_{-t}^{t} \Big[ -kB_{1}(t,s) + \ti{C}_{1}(t,s) - \frac{1}{k} \ti{D}_{1}(t,s) \Big] e^{i\gom(k)s} ds \notag \\
&= \int_{0}^{t} \Big[ -2kB_{1}(t,2\tau-t) + 2C_{1}(t,2\tau-t) + \frac{1}{2\gl} g_{0}(t) A_{1}(t,2\tau-t) \Big] e^{2i\gom(k)(\tau-t)} d\tau, \notag \\
&\hspace{5mm} - \frac{1}{k} \int_{0}^{t} \Big[ D_{1}(t,2\tau-t) - \frac{1}{2\gl} g_{0}(t) B_{2}(t,2\tau-t) - \frac{1}{4i \gl} g_{1}(t) A_{1}(t,2\tau-t) \Big] e^{2i\gom(k)(\tau-t)} d\tau, \label{Phi1-GLM}
\end{align*}
where we have used the change of variables $s \mapsto 2\tau-t$.  The definition of $B(t,k)$ (equation \eqref{spectral(A,B)}) then yields
\begin{subequations} \label{spectrals-A,B}
\begin{equation}
B(t,k) = 2k\h{B}_{1} - \big(2\h{C}_{1} + \frac{1}{2\gl} g_{0}(t) \h{A}_{1} \big) + \frac{1}{k} \big( 2\h{D}_{1} - \frac{1}{2\gl} g_{0}(t) \h{B}_{2} - \frac{1}{4i\gl} g_{1}(t) \h{A}_{1} \big),
\end{equation}
where
\begin{equation} \label{t-transform}
\h{N}(t,k) = \int_{0}^{t} N(t,2\tau-t) e^{2i\gom(k)\tau} d\tau.
\end{equation}
\end{subequations}
An analogous expression can also be written for $A(t,k)$.

\section{The generalized Dirichlet to Neumann map} \label{D2N}

As was noted earlier (see also Appendix A), given $q_{0}(x)$ and a ``proper''\footnote{``Proper'' here means those boundary conditions for which the initial-boundary value problem is at least linearly well posed.  For example, if $\gl=-1$ the boundary value problem with $q(0,t)$ given is well posed.  Similarly, if $\gl=1$ the boundary value problem with $\{ q(0,t),q_{x}(0,t) \}$ given is well posed.} subset of the boundary values $\{ g_{0}(t), g_{1}(t), g_{2}(t) \}$ we can characterize the unknown boundary values by the requirement that the spectral functions $\{A(t,k), B(t,k)\}$ satisfy the global relation \eqref{global-scalar}.  For the Riemann--Hilbert problem formulated in Theorem A.3 (see Appendix A below), the spectral functions $\{A(t,k), B(t,k)\}$ are needed in the domain
\begin{equation*}
D_{3} = \{ k \in \mathbb{C}, \mathrm{Im}\, k > 0 \cap \mathrm{Im}\, \gom(k) > 0 \}.
\end{equation*}
However, since equation \eqref{global-scalar} is valid for $\mathrm{Im}\, k \leq 0$ we must transform the global relation from the lower half $k$-plane into $D_3$.  We do this by using the invariance properties of the exponential $e^{2i\gom(k)t}$.  

For the case $\gl = -1$, the two non-trivial roots of the dispersion relation $\gom(p) = \gom(k)$ are such that $k \in D_{3} \Rightarrow p_{+}(k), p_{-}(k) \in D_{1}$ where $D_{1}$ is defined as
\begin{equation*}
D_{1} = \{ k \in \mathbb{C}, \mathrm{Im}\, k < 0 \cap \mathrm{Im}\, \gom(k) > 0 \}.
\end{equation*}
Evaluating the global relation \eqref{global-scalar} at $p_{+}$ and $p_{-}$ we thus find two equations for the spectral functions $A(t,k)$ and $B(t,k)$, which are valid for $k \in D_{3}$
\begin{subequations} \label{globals-KdV} 
\begin{align} 
a(p_{+})B(t,p_{+}) - b(p_{+})A(t,p_{+}) &= e^{2i\gom(k)t}c(t,p_{+}), \\
a(p_{-})B(t,p_{-}) - b(p_{-})A(t,p_{-}) &= e^{2i\gom(k)t}c(t,p_{-}), \quad k \in D_{3}.
\end{align}
\end{subequations}
These two equations give two independent relations coupling the known spectral functions $\{a(k), b(k)\}$ with the unknown spectral functions $\{A(t,k), B(t,k)\}$.  If we specify appropriate boundary conditions, for instance $q(0,t)=g_{0}(t)$, then we can use equations \eqref{globals-KdV} to determine the unknown boundary values $g_{1}(t)$ and $g_{2}(t)$ in terms of the given initial and boundary conditions.

\begin{figure}
\psfrag{a}{$D_{1}$}
\psfrag{b}{$D_{3}^{(1)}$}
\psfrag{c}{$D_{3}^{(2)}$}
\begin{center} 
\includegraphics{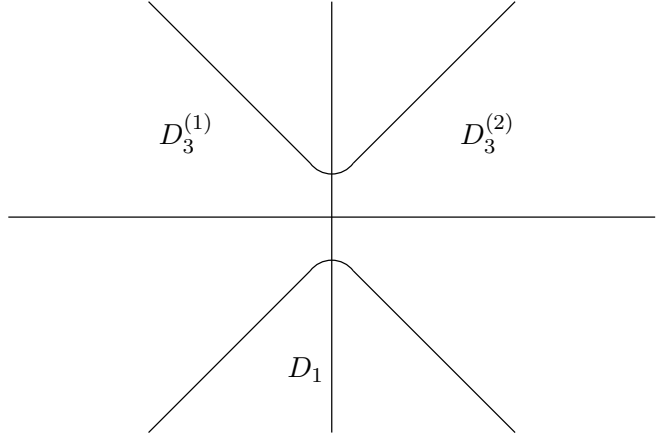}
\caption{The domains $D_{1}$, $D_{3}^{(1)}$, and $D_{3}^{(2)}$.} \label{D3}
\end{center} 
\end{figure}

An analogous situation also exists for the case $\gl = 1$.  In this case the two non-trivial roots of the dispersion relation $\gom(\nu) = \gom(k)$ have the following properties:
\begin{align*}
k \in D_{3}^{(1)} &\Longrightarrow \nu_{+}(k) \in D_{3}^{(2)}, \quad \nu_{-}(k) \in D_1, \\
k \in D_{3}^{(2)} &\Longrightarrow \nu_{+}(k) \in D_{1}, \quad \nu_{-}(k) \in D_{3}^{(1)}, 
\end{align*}
where $D_{3}^{(1)}$, $D_{3}^{(2)}$ are shown in Figure~\ref{D3}.
If we adopt the notation whereby $\nu(k) = \{ \nu_{+}(k), \nu_{-}(k) \}$ is defined such that $k \in D_{3} \Rightarrow \nu(k) \in D_{1}$, then from \eqref{global-scalar} we have 
\begin{equation} \label{global-KdV}
a(\nu)B(t,\nu) - b(\nu)A(t,\nu) = e^{2i\gom(k)t}c(t,\nu), \quad k \in D_{3}.
\end{equation}
This equation is a single relation coupling the known spectral functions $\{a(\nu), b(\nu)\}$ with the
unknown spectral functions $\{A(t,\nu), B(t,\nu)\}$.  If we specify appropriate boundary conditions, for instance $q(0,t) = g_{0}(t)$ and $q_{x}(0,t) = g_{1}(t)$, then we can use equation \eqref{global-KdV} to determine the unknown boundary value $g_{2}(t)$ in terms of the given initial and boundary conditions.

Before proceeding to solve equations \eqref{globals-KdV} and \eqref{global-KdV} for the unknown boundary values, we first introduce certain integral identities that will be required for our analysis:
\begin{subequations} \label{identities1}
\begin{align}
\int_{\pa{D}} \left[ \int_{0}^{t'} e^{2i\gom(k)(\tau-t')} K(t,\tau) d\tau \right] dk &= -\frac{\pi}{(3\gl)^{1/3}} \int_{0}^{t'} \frac{Ai(\gz)}{(t'-\tau)^{1/3}} K(t,\tau) d\tau, \\
\int_{\pa{D}} k \left[ \int_{0}^{t'} e^{2i\gom(k)(\tau-t')} K(t,\tau) d\tau \right] dk &= \frac{i\pi}{2(3\gl)^{2/3}} \int_{0}^{t'} \frac{Ai'(\gz)}{(t'-\tau)^{2/3}} K(t,\tau) d\tau,
\end{align}
\end{subequations}
where $Ai(\gz)$ is the solution of Airy's equation given in \eqref{Airy}, $K(t,\tau)$ is a smooth function of the arguments indicated,
\[
0 < t' < t,
\]
and the contour $\pa{D}$ is the oriented contour shown in Figure~\ref{boundary-D1*}.  In order to prove the first identity we first interchange the order of integration and then introduce new variables
\begin{equation*}
s = -2(3\gl)^{1/3}(t'-\tau)^{1/3}k, \quad \gz = (3\gl)^{-1/3}(t'-\tau)^{2/3}
\end{equation*}
to evaluate the $k$-integral.  The result then follows from equation \eqref{Airy}.  Similar considerations apply to prove the second identity above.

We will also make use of the following integral identities:
\begin{subequations} \label{identities2}
\begin{align}
\int_{\pa{D}} \gom'(k) \left[ \int_{0}^{t} e^{2i\gom(k)(\tau-t')} K(t,\tau) d\tau \right] dk &= \pi K(t,t') \label{FT1} \\
\int_{\pa{D}} \gom(k) \left[ \int_{0}^{t} e^{2i\gom(k)(\tau-t')} K(t,\tau) d\tau \right] dk &= -\frac{i\pi}{2(3\gl)^{1/3}} \int_{0}^{t'} \frac{Ai(\gz)}{(t'-\tau)^{1/3}} \frac{\pa{K}(t,\tau)}{\pa{\tau}} d\tau, \label{FT2} \\
\int_{\pa{D}} k\gom(k) \left[ \int_{0}^{t} e^{2i\gom(k)(\tau-t')} K(t,\tau) d\tau \right] dk &= \frac{\pi}{4(3\gl)^{2/3}} \int_{0}^{t'} \frac{Ai'(\gz)}{(t'-\tau)^{2/3}} \frac{\pa{K}(t,\tau)}{\pa{\tau}} d\tau. \label{FT3}
\end{align}
\end{subequations}
Identity \eqref{FT1} follows from the usual Fourier transform identity by mapping $\pa{D}$ to the real axis.  In order to derive expression \eqref{FT2} we first write the LHS as follows
\begin{equation*}
\int_{\pa{D}} \left[ \gom(k) \int_{0}^{t'} e^{2i\gom(k)(\tau-t')} K(t,\tau) d\tau - \frac{K(t,t')}{2i} \right] dk.
\end{equation*}
This ensures that the integral is well defined.  We next perform integration-by-parts on the $t$-integral to obtain
\begin{equation*}
\int_{\pa{D}} \frac{1}{2i} \left[ K(t,t') - \int_{0}^{t'} e^{2i\gom(k)(\tau-t')} \pa{K(t,\tau)}{\pa{\tau}} d\tau - K(t,t') \right] dk.
\end{equation*}
The two terms involving $K(t,t')$ cancel and, by interchanging the order of integration and using the identities in \eqref{identities1}, we can evaluate the $k$-integral to obtain the expression \eqref{FT2}.  Similar considerations apply to prove the third identity \eqref{FT3}.

\subsection{The solution of the global relation for $\gl = -1$}

We will now use the two global relations \eqref{globals-KdV} to derive equations \eqref{Isolutions}, i.e.~an explicit expression for the unknown boundary values $g_{1}(t)$ and $g_{2}(t)$.  Substituting into equations \eqref{globals-KdV} the expression for $B(t,k)$ given by \eqref{spectrals-A,B} and rearranging we find
\begin{multline*}
2\h{C}_{1} - \frac{1}{2} g_{0} \h{A}_{1} = 2(p_{+}+p_{-}) \h{B}_{1} - \frac{1}{p_{+}-p_{-}} \left( p_{+}\frac{b(p_{+})}{a(p_{+})}A(t,p_{+}) - p_{-}\frac{b(p_{-})}{a(p_{-})}A(t,p_{-}) \right) \\
- \frac{1}{p_{+}-p_{-}} e^{2i\gom(k)t} \left( p_{+}\frac{c(t,p_{+})}{a(p_{+})} - p_{-}\frac{c(t,p_{-})}{a(p_{-})} \right)
\end{multline*}
and
\begin{multline*}
2\h{D}_{1} + \frac{1}{2} g_{0} \h{B}_{2} + \frac{1}{4i} g_{1} \h{A}_{1} = 2p_{+}p_{-} \h{B}_{1} - \frac{p_{+}p_{-}}{p_{+}-p_{-}} \left( \frac{b(p_{+})}{a(p_{+})}A(t,p_{+}) - \frac{b(p_{-})}{a(p_{-})}A(t,p_{-}) \right) \\
- \frac{p_{+}p_{-}}{p_{+}-p_{-}} e^{2i\gom(k)t} \left( \frac{c(t,p_{+})}{a(p_{+})} - \frac{c(t,p_{-})}{a(p_{-})} \right),
\end{multline*}
where $\{ \h{A}_{1}, \h{B}_{1}, \h{B}_{2}, \h{C}_{1}, \h{D}_{1} \}$ are defined by equation \eqref{t-transform}.  We first assume that $a(k) \neq 0$ for $k \in \overline{D_{1}}$.  If we multiply these two equations by $\gom'(k) exp[-2i\gom(k)t']$ and integrate along $\pa{D}$, the contour shown in Figure~\ref{boundary-D1*}, then terms involving $c(t,p_{\pm})$ will vanish since $\gom'(k) c(t,p_{\pm})/a(p_{\pm})$ is analytic and of $O(1)$ for $\mathrm{Im}\, k \leq 0$, and the oscillatory term $\exp[2i\gom(k)(t-t')]$ is bounded in $D_{1}$.  We use the fact that $k\gom'(k)=(3\gom(k)-2)$ together with the integral identities \eqref{identities1} and \eqref{identities2} to obtain
\begin{multline*}
2\pi C_{1}(t,2t'-t) = \pi g_{0}(t) A_{1}(t,2t'-t) - \pi 3^{-1/3} \int_{0}^{t'} \frac{Ai(\gz)}{(t'-\tau)^{1/3}} \Big[ \frac{3i}{2} \frac{\pa{B}_{1}}{\pa{\tau}} - 2B_{1} \Big] (t,2\tau-t) d\tau \\
- \int_{\pa{D}} \gom'(k) e^{-2i\gom(k)t'} \frac{1}{p_{+} - p_{-}} \Big[ p_{+} \frac{b(p_{+})}{a(p_{+})} A(t,p_{+}) - p_{-} \frac{b(p_{-})}{a(p_{-})} A(t,p_{-}) \Big] dk, \\
\end{multline*}
and
\begin{multline*}
2\pi D_{1}(t,2t'-t) = -\frac{\pi}{2} g_{0}(t) B_{2}(t,2t'-t) + \frac{\pi}{4i} g_{1}(t) A_{1}(t,2t'-t) - \frac{c\pi}{2} B_{1}(t,2t'-t) \\
+ \frac{\pi}{2} 3^{-2/3} \int_{0}^{t'} \frac{Ai'(\gz)}{(t'-\tau)^{2/3}} \Big[ \frac{3}{2} \frac{\pa{B}_{1}}{\pa{\tau}} - 2iB_{1} \Big] (t,2\tau-t) d\tau \\
- \int_{\pa{D}} \gom'(k) e^{-2i\gom(k)t'} \frac{p_{+}p_{-}}{p_{+} - p_{-}} \Big[ \frac{b(p_{+})}{a(p_{+})} A(t,p_{+}) - \frac{b(p_{-})}{a(p_{-})} A(t,p_{-}) \Big] dk.
\end{multline*}
Letting $t' \to t$ in these expressions and using the identities for $B_{1}(t,t)$, $C_{1}(t,t)$ and $D_{1}(t,t)$ in \eqref{charac-value-BCs2+} we find \eqref{Isolutions}.

Finally, we note that if $a(k)$ does have zeros for $k \in \overline{D_{1}}$, then the integral involving $\gom'(k)c(t,k)\exp[2i\gom(k)(t-t')]/a(k)$ does not vanish, but rather it has a contribution due to the relevant residues.  The simplest way to compute this contribution is to deform the contour $\pa{D}$ to include the relevant zeros.

\subsection{The solution of the global relation for $\gl = 1$}

For the case of $\gl = 1$, we use the single global relation \eqref{global-KdV} to determine the unknown boundary value $g_{2}(t)$.  Substituting into equation \eqref{global-KdV} the expression for $B(t,k)$ given by \eqref{spectrals-A,B} and rearranging we find
\begin{equation*}
2\h{D}_{1} = \frac{1}{2} g_{0} \h{B}_{2} + \frac{1}{4i} g_{1} \h{A}_{1} - 2\nu^{2} \h{B}_{1} + \nu \big[ 2\h{C}_{1} + \frac{1}{2}g_{0} \h{A}_{1} \big] + \frac{b(\nu)}{a(\nu)} A(t,\nu) + e^{2i\gom(k)t}\frac{c(t,\nu)}{a(\nu)}.
\end{equation*}
We first assume that $a(\nu) \neq 0$ for $\nu \in \overline{D_{1}}$.  If we multiply this equation by $\gom'(\nu) exp[-2i\gom(k)t']$ and integrate along $\pa{D}_{1}$, then the final term on the RHS vanishes because $\gom'(\nu) c(t,\nu)/a(\nu)$ is analytic and of $O(1)$ for $\mathrm{Im}\, k \leq 0$, and the oscillatory term $\exp[2i\gom(k)(t-t')]$ is bounded in $D_{1}$.  Using the fact that $k\gom'(k)=(3\gom(k)-2)$ together with the integral identities \eqref{identities1} and \eqref{identities2} we find
\begin{multline*}
2\pi D_{1}(t,2t'-t) = \frac{\pi}{2} g_{0}(t) B_{2}(t,2t'-t) + \frac{\pi}{4i} g_{1}(t) A_{1}(t,2t'-t) \\
- \frac{\pi}{2} 3^{-2/3} \int_{0}^{t'} \frac{Ai'(\gz)}{(t'-\tau)^{2/3}} \Big[ \frac{3}{2}\frac{\pa{B}_{1}}{\pa{\tau}} - 2iB_{1} \Big] (t,2\tau-t) d\tau \\
- \pi 3^{-1/3} \int_{0}^{t'} \frac{Ai(\gz)}{(t'-\tau)^{1/3}} \Big[ 3i\frac{\pa{C}_{1}}{\pa{\tau}} - 4C_{1} + g_{0}(t) \Big( \frac{3i}{4}\frac{\pa{A}_{1}}{\pa{\tau}} - A_{1} \Big) \Big] (t,2\tau-t) d\tau \\
+ \int_{\pa{D}} k\gom'(k) e^{-2i\gom(k)t'} \frac{b(k)}{a(k)} A(t,k) dk.
\end{multline*}
Letting $t' \to t$ in this equation and using the identity for $D_{1}(t,t)$ in \eqref{charac-value-BCs2+} we find \eqref{IIg2}.

As in the case for $\gl = -1$, we note that if $a(k)$ does have zeros for $k \in \overline{D_{1}}$, then the integral involving $\gom'(k)c(t,k)\exp[2i\gom(k)(t-t')]/a(k)$ does not vanish.  The simplest way to compute the contribution from this term is to deform the contour $\pa{D}_{1}$ to include the relevant zeros.

\section{Conclusion} \label{Conclusion}

The generalized Dirichlet to Neumann map for integrable nonlinear evolution PDEs can be characterized by the requirement that the spectral functions $\{ a(k),b(k),A(t,k),B(t,k) \}$ satisfy the so-called global relation.  The Gelfand--Levitan--Marchenko representation of $\Phi^{(t)}(t,k)$ was used in \cite{BdMFS2003} and \cite{ASF2004dirich} to solve the global relation \emph{explicitly} for the nonlinear Schr\"odinger, the sine-Gordon, and the modified KdV equations (see also \cite{BdMFS2004}).  We have shown in this paper that the global relation can also be solved explicitly for both cases $\gl = \pm 1$ of the KdV equation.

The analysis of the KdV equation in the case $\gl = -1$ is more complicated than the analysis of the case $\gl = 1$ since in the former case the \emph{single} global relation \eqref{global-scalar} must be solved for \emph{two} unknown boundary values.  For economy of presentation we have concentrated on the case where $g_{0}(t)$ is prescribed as a boundary condition and, by using certain invariant transformations of the dispersion relation $\gom(k) = -4k^{3} + c k$, we have constructed explicit expressions for $g_{1}(t)$ and $g_{2}(t)$.  The analysis is similar for the case that, say $g_{1}(t)$, is prescribed.

The explicit solution of the global relation yields the unknown boundary values in terms of the given initial and boundary conditions and in terms of the kernel functions that appear in the Gelfand--Levitan--Marchenko representation of $\Phi^{(t)}(t,k)$.  These kernel functions in turn satisfy a system of nonlinear ODEs which are equivalent to a system of nonlinear Volterra integral equations for which the nonlinearity now appears in a simple explicit form.  The rigorous analysis of this system has not yet been carried out, however because the nonlinear integral equations are of Volterra type it should be straightforward to establish existence of the solution at least for small $t$ or for ``small'' boundary conditions \cite{ASF2004dirich}.  

We remark that the explicit characterization of the Dirichlet to Neumann map that we have presented here is a significant simplification on earlier attempts.  Furthermore, this development appears to be particularly useful in the context of numerical simulations of soliton equations.  Indeed, the result of \cite{ASF2004dirich} has been used in \cite{Z2006} for the implementation of the so-called method of nonreflecting boundary conditions for the numerical integration of the modified KdV equation.  Actually, it was in \cite{Z2006} that the formulae of \cite{ASF2004dirich} were simplified by considering certain $k$-integrals.  In this paper we have extended this analysis and applied it to the KdV equation.

Finally, we note that initial-boundary value problems for integrable nonlinear PDEs can also be analysed using PDE techniques \cite{BSZ2002}--\cite{CK2002}.  In particular we note the remarkable result of \cite{CK2002} where global well posedness is established.  We consider our approach as complementary to the above approach:  Having established \emph{a priori} global existence for $q(x,t)$ and therefore for $\{ q(0,t),q_{x}(0,t),q_{xx}(0,t) \}$, our results can be extended globally.  Therefore, our results imply an explicit characterization of the unknown boundary values which appears interesting both for analytical as well as computational considerations.  Also, we note the Riemann--Hilbert formalism reviewed in Appendix A together with the powerful Deift--Zhou method \cite{DZ1993} yield explicit results for the large $t$-asymptotics of the solution.  

\medskip \noindent
\textbf{Acknowledgement.}
This work was partially supported by the Department of Applied Mathematics and Theoretical Physics at the University of Cambridge, by Trinity College, Cambridge, and by the EPSRC.

\appendix

\section*{Appendix A Lax pair, spectral functions and the ``inverse'' problem}
\label{spectral-KdV}

In this appendix we summarize the main results for the KdV equation presented in \cite{ASF2002nonlin}.  Details and proofs can be found in \cite{ASF2002nonlin} and \cite{FIS2001}.

Equation \eqref{KdV} admits the Lax pair formulation
\begin{subequations} \label{KdV-Lax}
\begin{align}
\mu_{x} - ik[\gs_{3},\mu] &= Q(x,t,k) \mu, \tag{A.1a} \\
\mu_{t} + i\gom(k)[\gs_{3},\mu] &= P(x,t,k) \mu, \quad k \in \mathbb{C}, \tag{A.1b} 
\end{align}
\end{subequations}
where $\gom(k)$ is defined by \eqref{dispersion}, the eigenfunction $\mu(x,t,k)$ is a $2\times2$ matrix-valued function, and the functions $Q(x,t,k)$, $P(x,t,k)$ are defined by
\begin{align*}
Q(x,t,k) &= \frac{\gl}{2k}(\gs_{2}-i\gs_{3})q(x,t), \\
P(x,t,k) &= -2k\gs_{2}q(x,t) + \gs_{1}q_{x}(x,t) + \frac{1}{2k} (\gs_{2} - i\gs_{3}) \left(q_{xx}(x,t) - \gl c q(x,t) - 2\gl q^2(x,t) \right),
\end{align*}
with
\begin{equation*}
\gs_{1} = 
\begin{pmatrix}
0 & 1 \\
1 & 0
\end{pmatrix}, \quad
\gs_{2} = 
\begin{pmatrix}
0 & -i \\
i & 0
\end{pmatrix}, \quad
\gs_{3} = 
\begin{pmatrix}
1 & 0 \\
0 & -1
\end{pmatrix},
\end{equation*}
and $[,]$ denotes the usual matrix commutator
\[
[\gs_{3},A] = \gs_{3}A - A\gs_{3}.
\]

\subsection*{A.1 The ``direct'' spectral problem}

The spectral functions denoted by $\{ a(k),b(k),A(t,k),B(t,k) \}$, are defined in terms of the initial condition and the boundary values by the \emph{direct map}
\[
\{ q_{0}(x), g_{0}(t), g_{1}(t), g_{2}(t) \} \mapsto \{ a(k), b(k), A(t,k), B(t,k) \}.
\]
In order to set up the map $q_{0}(x) \mapsto \{ a(k), b(k) \}$, we first introduce the matrix-valued function $\Psi^{(x)}(x,k)$.  This function is defined in terms of $q_{0}(x)$ as the following matrix solution of the $x$-part of the associated Lax pair (A.1) evaluated at $t=0$:
\begin{subequations} \label{differential-a,b}
\begin{equation} \label{diff-psi-x}
\pa{}_{x}\Psi^{(x)}(x,k) - ik[\gs_{3},\Psi^{(x)}(x,k)] = Q_{0}(x,k)\Psi^{(x)}(x,k), \tag{A.2a}
\end{equation}
where
\begin{equation} \label{Q-0}
Q_{0}(x,k) = \frac{\gl}{2k}(\gs_{2}-i\gs_{3})q_{0}(x), \tag{A.2b}
\end{equation}
\begin{equation}
\lim_{x \to \infty} \Psi^{(x)}(x,k) = I, \quad 0 < x < \infty, \tag{A.2c} 
\end{equation}
\end{subequations}
and for the first column of $\Psi^{(x)}$ we take $\mathrm{Im}\, k \geq 0$, whereas for the second column we take $\mathrm{Im}\, k \leq 0$.

\medskip \noindent \textbf{Definition A.1.} {\it
The spectral functions $a(k)$ and $b(k)$ associated with the initial condition $q_{0}(x)$ are defined by
\begin{equation} \label{spectral-a,b}
a(k) = \psi_{2}(0,k), \quad b(k) = \psi_{1}(0,k), \quad \mathrm{Im}\, k \leq 0, \tag{A.3}
\end{equation}
where $\psi_{1}(x,k)$ and $\psi_{2}(x,k)$ are the following components of the matrix-valued function $\Psi^{(x)}(x,k)$ defined by equation (A.2),
\begin{equation} \label{Psi-x}
\Psi^{(x)}(x,k) =
\begin{pmatrix}
\overline{\psi_{2}(x,\bar{k})} & \psi_{1}(x,k) \\
\overline{\psi_{1}(x,\bar{k})} & \psi_{2}(x,k)
\end{pmatrix}. \tag{A.4}
\end{equation}
}

In order to set up the map $\{ g_{0}(t), g_{1}(t), g_{2}(t) \} \mapsto \{ A(t,k), B(t,k) \}$, we introduce the matrix-valued function $\Phi^{(t)}(t,k)$.  This function is defined in terms of $\{g_{0}(t), g_{1}(t), g_{2}(t) \}$ as the following solution of the $t$-part of the associated Lax pair (A.1) evaluated at $x=0$:
\begin{subequations} \label{differential-A,B}
\begin{equation} \label{ODE-A,B}
\pa{}_{t}\Phi^{(t)}(t,k) + i\gom(k) [\gs_{3},\Phi^{(t)}(t,k)] = P_{0}(t,k)\Phi^{(t)}(t,k), \tag{A.5a}
\end{equation}
where
\begin{equation} \label{P-0}
P_{0}(t,k) = -2\gs_{2} g_{0} k + \gs_{1} g_{1} + \frac{1}{2} \big( \gs_{2} - i\gs_{3} \big) \big(g_{2}(t) - \gl c g_{0}(t) - 2\gl g_{0}(t)^2 \big) \frac{1}{k}, \tag{A.5b}
\end{equation}
and
\begin{equation}
\Phi^{(t)}(0,k) = I, \tag{A.5c}
\end{equation}
\end{subequations}
and $k \in \mathbb{C}$.\footnote{If $t \to \infty$, we take $k$ such that $\mathrm{Im}\, \gom(k) \geq 0$ for the first column of $\Phi^{(t)}$, and $k$ such that $\mathrm{Im}\, \gom(k) \leq 0$ for the second column.}

\medskip \noindent \textbf{Definition A.2.} {\it
The spectral functions $A(t,k)$ and $B(t,k)$ associated with the boundary values $\{g_{0}(t), g_{1}(t), g_{2}(t) \}$ are defined by
\begin{equation} \label{spectral-A,B}
\begin{pmatrix}
-e^{-2i\gom(k)t}B(t,k) \\
\overline{A(t,\bar{k})}
\end{pmatrix} = 
\begin{pmatrix}
\Phi_{1}(t,k) \\
\Phi_{2}(t,k)
\end{pmatrix}, \tag{A.6}
\end{equation}
where $\Phi_{1}(t,k)$ and $\Phi_{2}(t,k)$ are the (12) and (22) components, respectively, of the matrix-valued function $\Phi^{(t)}(t,k)$ defined by (A.3).
}

\subsection*{A.2 The global relation}

Although the relationship between the initial condition and the boundary values of the solution of the KdV equation is complicated, this relationship takes a surprisingly simple form when expressed in terms of the corresponding spectral functions, see equation \eqref{global-scalar}.  In the case $0 < t \leq \infty$, equation \eqref{global-scalar} becomes
\begin{equation} \label{global-scalar2}
a(k)B(\infty,k) - b(k)A(\infty,k) = 0, \quad \mathrm{Im}\, k \leq 0 \cap \mathrm{Im}\, \gom(k) \geq 0. \tag{A.7}
\end{equation}
The derivation of the global relation can be found in \cite{ASF2002nonlin}.

\subsection*{A.3 The ``inverse'' spectral problem}

The \emph{inverse map}
\[
\{ a(k), e^{-2i\gom(k)t}b(k), A(T,k), e^{-2i\gom(k)t}B(T,k) \} \mapsto q(x,t),
\]
defines the function $q(x,t)$ by
\begin{equation} \label{RH-solution}
q(x,t) = \lim_{k \to \infty} 4k^{2} \left( M(x,t,k) \right)_{12}, \quad 0 < x < \infty, \quad 0 < t < T, \tag{A.8}
\end{equation}
where $M(x,t,k)$ satisfies the $2\times2$ matrix RH problem formulated in Theorem A.3, which we state below without proof.  For conciseness we give only the solitonless case.  The inclusion of solitons is a simple procedure, the details of which can be found in \cite{ASF2002nonlin}.  An improved procedure for including solitons is presented in \cite{BdMK2000}.

\medskip \noindent \textbf{Theorem A.3.} \label{theorem} \textbf{\cite{ASF2002nonlin}} 
{\it
Let $q_{0}(x) \in \mathcal{S}(\mathbb{R}^{+})$.  Define $\{a(k),b(k)\}$ by equation \eqref{spectral-a,b}.  Suppose that there exist smooth functions $\{ g_{0}(t), g_{1}(t), g_{2}(t) \}$ satisfying $\{ g_{j}(0)=\pa{}_{x}^{j}q_{0}(0) \}_{0}^{2}$, such that the functions $\{A(t,k),B(t,k)\}$, defined by equation \eqref{spectral-A,B}, satisfy the global relation \eqref{global-scalar}, where $c(t,k)$ is analytic in $\{ k \in \mathbb{C} \setminus 0 \,|\, \mathrm{Im}\, k < 0 \}$ and is of $O(1/k)$ as $k \to \infty$.\footnote{For the case $0 <t \leq \infty$, equation \eqref{global-scalar} is replaced by \eqref{global-scalar2} and we further assume that $g_{0}(t)$, $g_{1}(t)$, $g_{2}(t)$ belong to $\mathcal{S}(\mathbb{R}^{+})$.} 

Let $M(x,t,k)$ be a solution of the following $2\times2$ matrix RH
problem:
\begin{itemize}
\item
$M$ is sectionally meromorphic in $k \in \mathbb{C} \setminus \{ k
  \,|\, \mathrm{Im}\, \gom(k) = 0 \}$.

\item
Along the contours $\{ k \,|\, \mathrm{Im}\, \gom(k) = 0 \}$, $M$ satisfies
the following jump conditions 
\begin{equation} \label{RHjump}
M_{-}(x,t,k) = M_{+}(x,t,k)J(x,t,k), \quad \mathrm{Im}\, \gom(k) = 0, \tag{A.9}
\end{equation}
where the matrices $M_{\pm}(x,t,k)$ are the limit values of $M$ as $k$ approaches $\{ \mathrm{Im}\, \gom(k) = 0 \}$, and the jump matrix $J(x,t,k)$ is defined in terms of the spectral functions $a(k)$, $b(k)$, $A(T,k)$, and $B(T,k)$ as follows:
\begin{equation} \label{RH-J}
J(x,t,k) = \left\{
\begin{aligned}
&\begin{pmatrix}
1 & 0 \\
\overline{\Gamma(\bar{k})}e^{2i\gt} & 1
\end{pmatrix}, \quad k \in \overline{D_{1}} \cap \overline{D_{2}} & \\
&\begin{pmatrix}
1 & -\Gamma(k)e^{-2i\gt} \\
0 & 1
\end{pmatrix} 
\begin{pmatrix}
1 - |\gga(k)|^2 & \gga(k)e^{-2i\gt} \\
-\overline{\gga(\bar{k})} e^{2i\gt} & 1
\end{pmatrix} \\
& \times
\begin{pmatrix}
1 & 0 \\
\overline{\Gamma(\bar{k})}e^{2i\gt} & 1
\end{pmatrix}, \quad k \in \overline{D_{2}} \cap \overline{D_{3}} \\
&\begin{pmatrix}
1 & -\Gamma(k)e^{-2i\gt} \\
0 & 1
\end{pmatrix}, \quad k \in \overline{D_{3}} \cap \overline{D_{4}},
\end{aligned} \right.  \tag{A.10}
\end{equation}
where
\begin{equation}
\gt(x,t,k) = -kx + \gom(k)t; \tag{A.11}
\end{equation}
the functions $\gga(k)$, $\Gamma(k)$ are given by
\begin{subequations} \label{RH-gammas}
\begin{align}
\gga(k) &= \frac{b(k)}{\overline{a(\bar{k})}},& &k \in \overline{D_{2}} \cap \overline{D_{3}},&  \tag{A.12a} \\
\Gamma(k) &= \frac{B(T,k)}{\overline{a(\bar{k})}\,\overline{d(\bar{k})}},& &k \in \overline{D_{2}} \cap \overline{D_{3}}, \quad \overline{D_{3}} \cap \overline{D_{4}},&  \tag{A.12b} \label{RH-Gamma}
\end{align}
with
\begin{equation}  \label{d(k)}
d(k) = a(k)\overline{A(T,\bar{k})} - b(k)\overline{B(T,\bar{k})}, \quad k \in D_{2}; \tag{A.12c}
\end{equation}
\end{subequations}
and the domains $D_{1}$--$D_{4}$ are defined as:
\begin{equation} \label{domains-KdV}
\begin{split}
D_1 &= \{k \in \mathbb{C}, \mathrm{Im}\; k < 0 \cap \mathrm{Im}\;
\gom(k) > 0 \}, \\
D_2 &= \{k \in \mathbb{C}, \mathrm{Im}\; k < 0 \cap \mathrm{Im}\;
\gom(k) < 0 \}, \\
D_3 &= \{k \in \mathbb{C}, \mathrm{Im}\; k > 0 \cap \mathrm{Im}\;
\gom(k) > 0 \}, \\
D_4 &= \{k \in \mathbb{C}, \mathrm{Im}\; k > 0 \cap \mathrm{Im}\;
\gom(k) < 0 \}.
\end{split} \tag{A.13}
\end{equation}

\item
\begin{equation} \label{RH-boundary}
M(x,t,k) = I + O\left(\frac{1}{k}\right), \quad k \to \infty. \tag{A.14}
\end{equation}

\item
$M(x,t,k)$ has a pole at $k=0$ satisfying
\[
M(x,t,k) \sim \frac{if(x,t)}{k}
\begin{pmatrix}
0 & 1 \\
0 & -1
\end{pmatrix}, \quad k \to 0.
\]

\end{itemize}
Then we have the following:
\begin{enumerate}
\item[(1)]
$M(x,t,k)$ exists and is unique.

\item[(2)]
$q(x,t)$ defined in terms of $M(x,t,k)$ by equation \eqref{RH-solution} satisfies the KdV equation \eqref{KdV}.

\item[(3)]
$q(x,t)$ satisfies the initial condition $q(x,0) = q_{0}(x)$, and furthermore, it has the following boundary values:
\begin{equation*}
q(0,t) = g_{0}(t), \quad q_{x}(0,t) = g_{1}(t), \quad q_{xx}(0,t) = g_{2}(t).
\end{equation*}
\end{enumerate}
}

\medskip \noindent
\textbf{Proof.}  Details of the proof of this theorem can be found in \cite{ASF2002nonlin}.

\section*{Appendix B The linear limit} \label{linear}

In this appendix we give the generalized Dirichlet to Neumann map for the linear version of equation \eqref{KdV}.  We also prove that, in the linear limit, our construction of the Dirichlet to Neumann map for the KdV equation in Section~\ref{results} reduces to the corresponding map in the linear case.

In the approximation of small $q_{0}$, $g_{0}$, $g_{1}$, $g_{2}$ (or small $q_{0}$ and small $t$) then, from Proposition~\ref{proposition}, we find $B_{1}(t,s) = B_{1}(t+s)$ and thus $B_{1}(t,t) = B_{1}(2t) = -ig_{0}(t)$.  Hence
\begin{equation*}
B_{1}(t,2\tau-t) = B_{1}(2\tau) = -ig_{0}(\tau).
\end{equation*}
Similar considerations apply for the other kernel functions and so we have
\begin{align*}
B_{1}(t,2\tau-t) &\approx -ig_{0}(\tau),& \quad A_{1}(t,2\tau-t) &\approx 0, \\
C_{1}(t,2\tau-t) &\approx \frac{1}{2} g_{1}(\tau),& \quad B_{2}(t,2\tau-t) &\approx 0, \\
D_{1}(t,2\tau-t) &\approx \frac{i}{4} \big[g_{2}(\tau) - \gl c g_{0}(\tau) \big],& \quad C_{2}(t,2\tau-t) &\approx 0.
\end{align*}
and, from \eqref{a,b},
\begin{equation*}
a(k) \sim 1, \quad b(k) \sim -\frac{1}{2ik} \int_{0}^{\infty} e^{-2ikx} q_{0}(x) dx.
\end{equation*}

\subsection*{B.1 The linear limit for KdVI $(\gl = -1)$} \label{linearKdVI}
Taking the linear limit of \eqref{Isolutions} and substituting the above expressions, we find
\begin{multline*}
g_{1}(t) \approx \frac{2}{\pi} \int_{D} ik\gom'(k) \Big[ \int_{0}^{t} e^{2i\gom(k)(\tau-t)} g_{0}(\tau) d\tau \Big] dk \\
- \frac{1}{\pi} \int_{\pa{D}} \gom'(k)e^{-2i\gom(k)t} \frac{1}{p_{+} - p_{-}} \Big[ p_{+} \h{q}_{0}(p_{+}) - p_{-} \h{q}_{0}(p_{-}) \Big] dk,
\end{multline*}
and
\begin{multline*}
g_{2}(t) \approx \frac{4}{\pi} \int_{D} k^{2}\gom'(k) \Big[ \int_{0}^{t} e^{2i\gom(k)(\tau-t)} g_{0}(\tau) d\tau \Big] dk \\
+ \frac{2i}{\pi} \int_{\pa{D}} \gom'(k)e^{-2i\gom(k)t} \frac{p_{+}p_{-}}{p_{+} - p_{-}} \Big[ \h{q}_{0}(p_{+}) - \h{q}_{0}(p_{-}) \Big] dk,
\end{multline*}
where
\begin{equation*}
\h{q}_{0}(k) = \int_{0}^{\infty} e^{-2ikx} q_{0}(x) dx.
\end{equation*}
By using the identities \eqref{identities1} and \eqref{identities2} we see that these are the same formulae we get by solving the global relation  associated with equation \eqref{KdVlinear} with $\gl = -1$, see \cite{ASF2002new}.

\subsection*{B.2 The linear limit for KdVII $(\gl = 1)$} \label{linearKdVII}
Taking the linear limit of \eqref{IIg2} and substituting the above expressions, we find
\begin{multline*}
g_{2}(t) \approx c g_{0}(t) + \frac{8}{\pi} \int_{D} k^{2}\gom'(k) \Big[ \int_{0}^{t} e^{2i\gom(k)(\tau-t)} g_{0}(\tau) d\tau \Big] dk \\
-\frac{4i}{\pi} \int_{D} k\gom'(k) \Big[ \int_{0}^{t} e^{2i\gom(k)(\tau-t)} g_{1}(\tau) d\tau \Big] dk \\
- \frac{4i}{\pi} \int_{\pa{D}} \gom'(k)e^{-2i\gom(k)t} \h{q}_{0}(k)dk,
\end{multline*}
where
\begin{equation*}
\h{q}_{0}(k) = \int_{0}^{\infty} e^{-2ikx} q_{0}(x) dx.
\end{equation*}
By using the identities \eqref{identities1} and \eqref{identities2} we see that these are the same formulae we get by solving the global relation  associated with equation \eqref{KdVlinear} with $\gl = 1$, see \cite{ASF2002new}.

\end{document}